\documentclass{aa}

\usepackage{psfig}

\let\a=\alpha
\let\l=\lambda

\def\ion#1#2{{\rm #1}\,{\sc #2}}

\usepackage{amssymb}

\begin{document}



\title{UVES observations of a damped Ly$\a$ system at $z_{\rm abs} = 4.466$ towards the quasar 
APM BR J0307$-$4945\thanks{Based on public data released from UVES Commissioning at the 
VLT/Kueyen telescope, ESO, Paranal, Chile.}}

\author{M. Dessauges-Zavadsky \inst1$^,$\inst2, S. D'Odorico \inst1, R.G. McMahon\inst3, 
P. Molaro \inst4, C. Ledoux \inst1, C. P\'{e}roux \inst3 \and \\
L.J. Storrie-Lombardi \inst5}

\institute{
European Southern Observatory, Karl-Schwarzschildstr. 2, 85748 Garching bei M\" unchen, Germany
\and
Observatoire de Gen\`eve, 1290 Sauverny, Switzerland
\and
Institute of Astronomy, Madingley Road, Cambridge CB3 0HA, England
\and
Osservatorio Astronomico di Trieste, Via G.B. Tiepolo 11, 34131 Trieste, Italy
\and
SIRTF Science Center, California Institute of Technology, MS 100-22, Pasadena, CA, USA}

\offprints{mdessaug@eso.org}


\date{Received / Accepted}

\authorrunning{M. Dessauges-Zavadsky et al.}

\titlerunning{UVES observations of a damped Ly$\a$ system at $z_{\rm abs} = 4.466$}

\maketitle


\begin{abstract}
We present the first high-resolution (6.2 to 7.7 km s$^{-1}$ FWHM) spectra of the APM BR J0307$-$4945 quasar at 
$z_{\rm em} = 4.73$ obtained with UVES on the 8.2m VLT Kueyen telescope. We focus our analysis on a damped Ly$\a$ 
(DLA) system at $z_{\rm abs} = 4.466$ with a neutral hydrogen column density $N$(\ion{H}{i}) $= (4.68\pm 0.97)\cdot10^{20}$ 
cm$^{-2}$. It is the most distant DLA system known to the present date, observed when the age of the universe was only 
1.3 Gyr. It shows complex low- and high-ionization line profiles spanning $\approx 240$ and 300 km s$^{-1}$ in velocity 
space respectively. We derive accurate abundances for N, O, Al, Si and Fe, and place a lower limit on C and an upper limit 
on Ni: [N/H] $=-3.07\pm 0.15$, [O/H] $=-1.63\pm 0.19$, [Al/H] $=-1.79\pm 0.11$, [Si/H] $=-1.54\pm 0.11$, 
[Fe/H] $=-1.97\pm 0.19$, [C/H] $>-1.63$ and [Ni/H] $<-2.35$. The derived high metallicity, $\sim 1/90$ solar, shows that 
this very young absorber ($\leq 1.3$ Gyr) has already experienced a significant metal enrichment. The [O/Si] ratio is 
nearly solar suggesting a limited amount of dust, the relative [Si,O/Fe] abundance ratios show a similar enhancement as 
observed in the Milky Way stars with comparable metallicities, and the [N/O] ratio is very low. All these results point 
to an enrichment pattern dominated by Type II supernovae which suggests a Milky Way type evolutionary model.
\end{abstract}



\section{Introduction}

For two decades now bright quasars (hereafter QSOs) have been used to investigate the distribution and physical properties 
of gas in their lines of sight up to their emission redshift. The information is extracted from the study of absorption 
lines originating in the gas (the Hydrogen Lyman series lines and resonance lines of various metal ions) as detected in the 
high dispersion ($10000 < R < 50000$) spectra of the QSOs.


\begin{table*}[!]
\begin{center}
\caption{Journal of observations} 
\label{journal}
\begin{tabular}{c c c c c c c c c}
\hline\hline
Number  & Date & Central    & Range & Resolution    & CD    & Slit & Seeing & Exposure 
\\
of      &      & wavelength &       & $\l/\Delta\l_{\scriptsize{\textrm{instr.}
}}$ & grating & width & & time\\
spectra &      & [\AA]      & [\AA] &               &       & ["]  & ["] &  [s]      
\smallskip
\\      
\hline
1 & 13/10/99 & 6000 & 5000--5960 & 39020 & \#3 & 1.1 & 0.8 & 4400           \\
  &          &      & 6120--7000 & 38980 &     &     &     &               \\
3 & 8/10/99  & 7600 & 5800--7500 & 43880 & \#4 & 1.0 & 0.7--0.9 & 3300/3000/3300 \\
  &          &      & 7710--9510 & 43550 &     &     &     &                \\
3 & 14/12/99 & 8000 & 6120--7960 & 48020 & \#4 & 0.9 & 0.8 & 4000/4000/4300 \\
  &          &      & 8100--9900 & 46090 &     &     &     &               \\
\hline
\end{tabular}
\begin{minipage}{160mm}
\smallskip
Mode: mosaic of the RED arm.
\end{minipage}
\end{center}
\end{table*}


A special class of absorption systems is represented by the damped Ly$\a$ systems. DLA systems are defined as absorption 
systems with a \ion{H}{i} column density $\geq 2\cdot 10^{20}$ cm$^{-2}$ which gives origin to the characteristic damped 
profile in the Ly$\a$ absorption. There is no general consensus on the nature of the galaxies associated with the DLA systems. 
The identification of all DLA systems with progenitors of present-day spirals as proposed originally by Wolfe et al. (1986) 
is not fully supported by the observations of optical counterparts at low redshifts which show a variety of morphological 
types (e.g. Le Brun et al. 1997) and by the relative metal abundances. 

Whatever their nature, DLA systems appear to be associated with the bulk of neutral hydrogen in the universe at high redshift
(Wolfe et al. 1995, Lanzetta et al. 1995, Storrie-Lombardi et al. 1996) and remain a unique way to trace accurately chemical 
abundances over a wide interval of redshifts, from 0.5 to 4.5, setting important constraints on theories of galaxy formation 
and evolution.

In the last decade, this type of research has especially benefitted from the use of the HIRES spectrograph at the Keck 
telescope. HIRES observations have produced a large number of accurate abundance measurements in DLA systems (Lu et al. 
1996,1997, Prochaska \& Wolfe 1999) which together with more complete DLA surveys (Rao \& Turnshek 2000, 
Storrie-Lombardi \& Wolfe 2000) have been used to explore both the redshift evolution of comoving mass density and 
of the metal content (see for a recent review Pettini 2000). It is unclear whether any of these quantities is significantly 
evolving from the early universe ($\sim 10$\% of the present age) to the present epoch, in contradiction with 
the simple picture describing a progressive conversion of gas into stars and an increase in the mean metallicity. Indeed, 
DLA systems at all redshifts might simply represent galaxies ``caught in the act'' of assembling a large 
amount of gas before a major episode of star formation. 

Whatever the favorite interpretation, it is clear that the sample of DLA systems with high-quality spectroscopic data 
(now summing up to about 70 objects) has to be substantially enlarged, to increase the statistical weight of the results 
at all redshifts and in particular at the low and very high redshift ends. The number of well studied systems at high 
redshifts is particularly low. The recent overview of abundance determinations by Prochaska \& Wolfe (2000) includes
13 objects at $z_{\rm abs} \geq 3$ and 5 objects at $z_{\rm abs} \geq 4$ only.

In October 1999, ESO installed at the second 8.2m VLT telescope (Kueyen) its high-resolution spectrograph, UVES, 
which compares well in efficiency and resolution with HIRES (D'Odorico et al. 2000). Taking advantage of its superior 
near-infrared efficiency, UVES has already been used to secure the first Zn abundance in a DLA system at $z_{\rm abs} > 3$ 
so far (Molaro et al. 2000). 

In this paper we report UVES observations of the highest redshift ($z_{\rm abs} = 4.466$) DLA system known to 
the present date. The quality of the data is such that a detailed analysis of the metal content of the absorbing gas has 
been possible and from this the evolutionary history of the associated galaxy has been reconstructed. While the statistical 
weight of the results on a single DLA system is obviously limited, with this study we demonstrate the possibility to 
investigate the star formation history of galaxies well beyond $z = 4$ from observations of relatively faint QSOs. 
The observations were obtained during the Commissioning of the instrument. They are part of the public set of UVES 
Commissioning data available through the ESO VLT archive.

In Sect.~2 we briefly review the observations and data reduction, and we provide a list of all metal systems we have 
identified in the line of sight to APM BR J0307$-$4945 redwards of the Ly$\a$ emission. We report the column density 
measurements of different ions of the DLA system at $z_{\rm abs} = 4.466$ in Sect.~3. Sect.~4 discusses the 
abundances and Sect.~5 the kinematics of the DLA gas. The conclusions are summarized in Sect.~6.


\section{Data}

\subsection{Observations}

APM BR J0307$-$4945 is a newly discovered quasar at $z_{\rm em}=4.73$ (R = 18.8) from the second APM color survey for 
$z > 4$ QSOs (Storrie-Lombardi et al. 2001). The damped absorber at $z_{\rm abs} = 4.466$ was so prominent that it was
identifiable in the low resolution (FWHM $\sim 10$ \AA) quasar discovery spectrum. On the basis of medium resolution 
spectroscopy (FWHM $\sim 5$ \AA) the DLA column density was estimated to be $\log N$(\ion{H}{i}) $= 20.8$ (P\'eroux 
et al. 2001 and McMahon et al. 2001, in preparation). 

The observations of APM BR J0307$-$4945 presented here were obtained during the first and second Commissioning of the 
Ultraviolet-Visual Echelle Spectrograph (UVES) on the Nasmyth focus of the VLT 8.2m Kueyen telescope at Paranal, 
in October and December 1999. A journal of observation dates, wavelength coverages, resolutions and exposure times of 
the data is presented in Table~\ref{journal}. Seven spectra, covering in total the spectral range from 5000 to 
9900 \AA, were obtained by using the cross-disperser gratings \#3 and \#4 of the spectrograph red arm. Details on the 
instrument can be found in Dekker et al. (2000) and in the UVES User Manual available at 
http://www.eso.org/instruments/uves/userman/. The full width at half maximum of the instrument profile, 
$\Delta \l_{\scriptsize{\textrm{instr.}}}$, was measured from the emission lines of the Thorium-Argon lamp and 
the resulting resolving powers, $\textrm{R} = \l/\Delta \l_{\scriptsize{\textrm{instr.}}}$, vary 
between 38980 and 48020 (median values) depending on the spectra (see Table~\ref{journal}) and correspond to 
a velocity resolution of 6.2 to 7.7 km s$^{-1}$ FWHM.

\subsection{Data reduction}

The data reduction was performed on each of the seven spectra separately using the UVES data reduction pipeline 
implemented in the ESO MIDAS package. The pipeline reduction is based on the following steps: order definition using 
a special order definition frame (order width $\simeq 5$ pixels); wavelength calibration; order extraction using 
the ``optimal'' extraction method (the orders are extracted by calculating a weighted sum of pixel values across 
the profile of the object); extraction of weighted flat-field; flat-fielding; and sky subtraction. Typical RMS errors 
in the wavelength calibrations are $\leq 5.5$ m\AA.

The observed wavelength scale was transformed into vacuum, heliocentric scale. The spectra were normalized by using 
a spline to smoothly connect the regions free from absorption features. The continuum for the Ly$\a$ forest region was 
fitted by using small regions deemed to be free of absorptions and by interpolating between these regions with a spline. 
Finally, the normalized spectra were added together using their S/N as weights. The final spectrum reaches a 
signal-to-noise of $20\leq \textrm{S/N}\leq 40$.

\subsection{Absorption line identification}

We have proceeded with the absorption-line identification by first trying to find all \ion{C}{iv}, 
\ion{Si}{iv} and \ion{Mg}{ii} doublets. Once we had composed a list of redshifts for the metal-line systems, we 
attempted to match the remaining absorption features with the strongest metal-line transitions. Finally, we compared 
in velocity space the line profiles of individual transitions in the determined systems for conclusive identification. 
The discrimination of telluric absorption lines from extragalactic absorption has been possible thanks to the spectrum of 
a standard star covering the spectral ranges from 6120 to 7960 \AA\ and from 8100 to 9900 \AA, and the comparison of 
spectra taken at different epochs.

Table~5\footnote{Available in electronic form at http://cdsweb.u-strasbg.fr} lists the wavelengths and equivalent 
widths for all absorption-line features (except the telluric lines) redwards of the Ly$\a$ emission that exceed the 
4$\sigma$ limit in equivalent width. The transition names and approximate redshifts are specified for the features we 
successfully identified. About 80\% of the absorption lines redwards of the Ly$\a$ emission have been identified, 
corresponding to 13 metal-line systems. In this work we will restrict our analysis to the damped Ly$\a$ system at 
$z_{\rm abs}=4.466$.


\section{Ionic column densities}\label{column-densities}

We present here the measurements of ionic column densities obtained for the damped Ly$\a$ system at 
$z_{\rm abs} = 4.466$. They have been derived by fitting theoretical Voigt profiles to the observed absorption
lines via a $\chi ^2$ minimization. The fits were performed using the FITLYMAN package included in MIDAS 
(Fontana \& Ballester 1995). During the fitting procedure the theoretical profiles were convolved with the instrumental 
point spread function modeled from the analysis of the emission lines of the arcs. The FITLYMAN routines determine 
for each absorption component the redshift, the column density, the turbulent broadening parameter $b$ and the fit 
errors for each of these quantities. The atomic data are taken from the compilation of Morton (1991) and from the 
updated values of Spitzer \& Fitzpatrick (1993) for silicon, Cardelli \& Savage (1995) for iron and Zsarg\' o \& Federman 
(1998) for nickel, as specified in Table~\ref{col-den}.


\begin{table*}[!]
\begin{center}
\caption{Fits for the DLA system at $z_{\rm abs} = 4.466$ - Low-ions} 
\label{low-ion-tab}
\begin{tabular}{l c c l c | l c c l c}
\hline\hline
Comp & $z_{\rm abs}$ & $b$    & Ident & $\log N$    & Comp & $z_{\rm abs}$ & $b$    & Ident & $\log N$  \\
     &               & [km s$^{-1}$]  &       & [cm$^{-2}$] &      &           & [km s$^{-1}$] &       & [cm$^{-2}$]  
\smallskip
\\     
\hline
1... & 4.464386 & 13.5 $\pm$ 0.9& \ion{O}{i}   & 13.12 $\pm$ 0.08 &7...  & 4.466582 & 9.4 $\pm$ 3.5& \ion{O}{i}   & 14.64 $\pm$ 0.70 \\
     &        &                 & \ion{C}{ii}  & 13.58 $\pm$ 0.02 &      &       &                 & \ion{C}{ii}  & $>$ 13.95 \\
     &        &                 & \ion{Si}{ii} & 12.84 $\pm$ 0.11 &      &       &                 & \ion{Si}{ii} & 13.61 $\pm$ 0.04\\
     &        &                 & \ion{Fe}{ii} & 13.00 $\pm$ 0.21 &      &       &                 & \ion{Fe}{ii} & 13.35 $\pm$ 0.25\\
     &        &                 & \ion{Al}{ii} & 11.85 $\pm$ 0.07 &      &       &                 & \ion{Al}{ii} & 11.59 $\pm$ 0.29\\
     &        &                 & \ion{N}{i}   & ...              &      &       &                 & \ion{N}{i}   & 12.68 $\pm$ 0.30\\
2... & 4.464686 & 4.6 $\pm$ 0.2 & \ion{O}{i}   & 13.59 $\pm$ 0.02 &8...  & 4.466844 & 6.5 $\pm$ 1.3& \ion{O}{i}   & 14.61 $\pm$ 0.35\\
     &        &                 & \ion{C}{ii}  & 13.76 $\pm$ 0.04 &      &       &                 & \ion{C}{ii}  & $>$ 14.23 \\
     &        &                 & \ion{Si}{ii} & 13.17 $\pm$ 0.09 &      &       &                 & \ion{Si}{ii} & 13.54 $\pm$ 0.02\\
     &        &                 & \ion{Fe}{ii} & 12.80 $\pm$ 0.23 &      &       &                 & \ion{Fe}{ii} & 12.69 $\pm$ 0.60\\
     &        &                 & \ion{Al}{ii} & 11.93 $\pm$ 0.06 &      &       &                 & \ion{Al}{ii} & 12.15 $\pm$ 0.11\\
     &        &                 & \ion{N}{i}   & ...              &      &       &                 & \ion{N}{i}   & 12.70 $\pm$ 0.12\\     
3... & 4.465225 & 7.2 $\pm$ 2.1 & \ion{O}{i}   & 12.37 $\pm$ 0.35 &9...  & 4.467511 & 11.7 $\pm$ 1.6& \ion{O}{i}  & 13.49 $\pm$ 0.04\\
     &        &                 & \ion{C}{ii}  & 13.13 $\pm$ 0.04 &      &       &                 & \ion{C}{ii}  & 13.25 $\pm$ 0.03\\
     &        &                 & \ion{Si}{ii} & 12.57 $\pm$ 0.10 &      &       &                 & \ion{Si}{ii} & 12.84 $\pm$ 0.07\\
     &        &                 & \ion{Fe}{ii} & 12.72 $\pm$ 0.30 &      &       &                 & \ion{Fe}{ii} & 12.57 $\pm$ 0.44\\
     &        &                 & \ion{Al}{ii} & 11.69 $\pm$ 0.09 &      &       &                 & \ion{Al}{ii} & 11.77 $\pm$ 0.08\\
     &        &                 & \ion{N}{i}   & ...              &      &       &                 & \ion{N}{i}   & ...\\
4... & 4.465510 & 6.2 $\pm$ 0.2 & \ion{O}{i}   & 14.02 $\pm$ 0.01 &10... & 4.467863 & 4.9 $\pm$ 0.8& \ion{O}{i}   & 15.33 $\pm$ 0.12\\
     &        &                 & \ion{C}{ii}  & 14.48 $\pm$ 0.04 &      &       &                 & \ion{C}{ii}  & $>$ 14.00 \\
     &        &                 & \ion{Si}{ii} & 13.67 $\pm$ 0.01 &      &       &                 & \ion{Si}{ii} & 13.57 $\pm$ 0.04\\
     &        &                 & \ion{Fe}{ii} & 13.11 $\pm$ 0.13 &      &       &                 & \ion{Fe}{ii} & 13.20 $\pm$ 0.13\\
     &        &                 & \ion{Al}{ii} & 12.45 $\pm$ 0.04 &      &       &                 & \ion{Al}{ii} & 12.05 $\pm$ 0.11\\
     &        &                 & \ion{N}{i}   & ...              &      &       &                 & \ion{N}{i}   & 12.84 $\pm$ 0.12\\     
5... & 4.465960 & 8.2 $\pm$ 0.5 & \ion{O}{i}   & 13.73 $\pm$ 0.02 &11... & 4.468005 & 9.6 $\pm$ 0.3& \ion{O}{i}   & 15.67 $\pm$ 0.14\\
     &        &                 & \ion{C}{ii}  & $>$ 14.43        &      &       &                 & \ion{C}{ii}  & $>$ 15.42 \\      
     &        &                 & \ion{Si}{ii} & 13.66 $\pm$ 0.02 &      &       &                 & \ion{Si}{ii} & 14.14 $\pm$ 0.02\\
     &        &                 & \ion{Fe}{ii} & 13.08 $\pm$ 0.14 &      &       &                 & \ion{Fe}{ii} & 13.69 $\pm$ 0.05\\
     &        &                 & \ion{Al}{ii} & 12.65 $\pm$ 0.04 &      &       &                 & \ion{Al}{ii} & 12.61 $\pm$ 0.04\\
     &        &                 & \ion{N}{i}   & ...              &      &       &                 & \ion{N}{i}   & 13.30 $\pm$ 0.08\\
6... & 4.466417 & 14.4 $\pm$ 2.2& \ion{O}{i}   & 13.97 $\pm$ 0.13 &12... & 4.468343 & 4.4 $\pm$ 0.5& \ion{O}{i}   & 14.17 $\pm$ 0.04\\
     &        &                 & \ion{C}{ii}  & $>$ 14.56        &      &       &                 & \ion{C}{ii}  & 13.16 $\pm$ 0.12\\
     &        &                 & \ion{Si}{ii} & 13.97 $\pm$ 0.22 &      &       &                 & \ion{Si}{ii} & 12.74 $\pm$ 0.06\\
     &        &                 & \ion{Fe}{ii} & 13.21 $\pm$ 0.29 &      &       &                 & \ion{Fe}{ii} & 12.64 $\pm$ 0.26\\
     &        &                 & \ion{Al}{ii} & 12.77 $\pm$ 0.03 &      &       &                 & \ion{Al}{ii} & 11.48 $\pm$ 0.13\\
     &        &                 & \ion{N}{i}   & ...              &      &       &                 & \ion{N}{i}   & ...\\
\hline 
\end{tabular}
\end{center}
\end{table*}


\begin{figure*}[!]
\begin{center}
\mbox{\psfig{figure=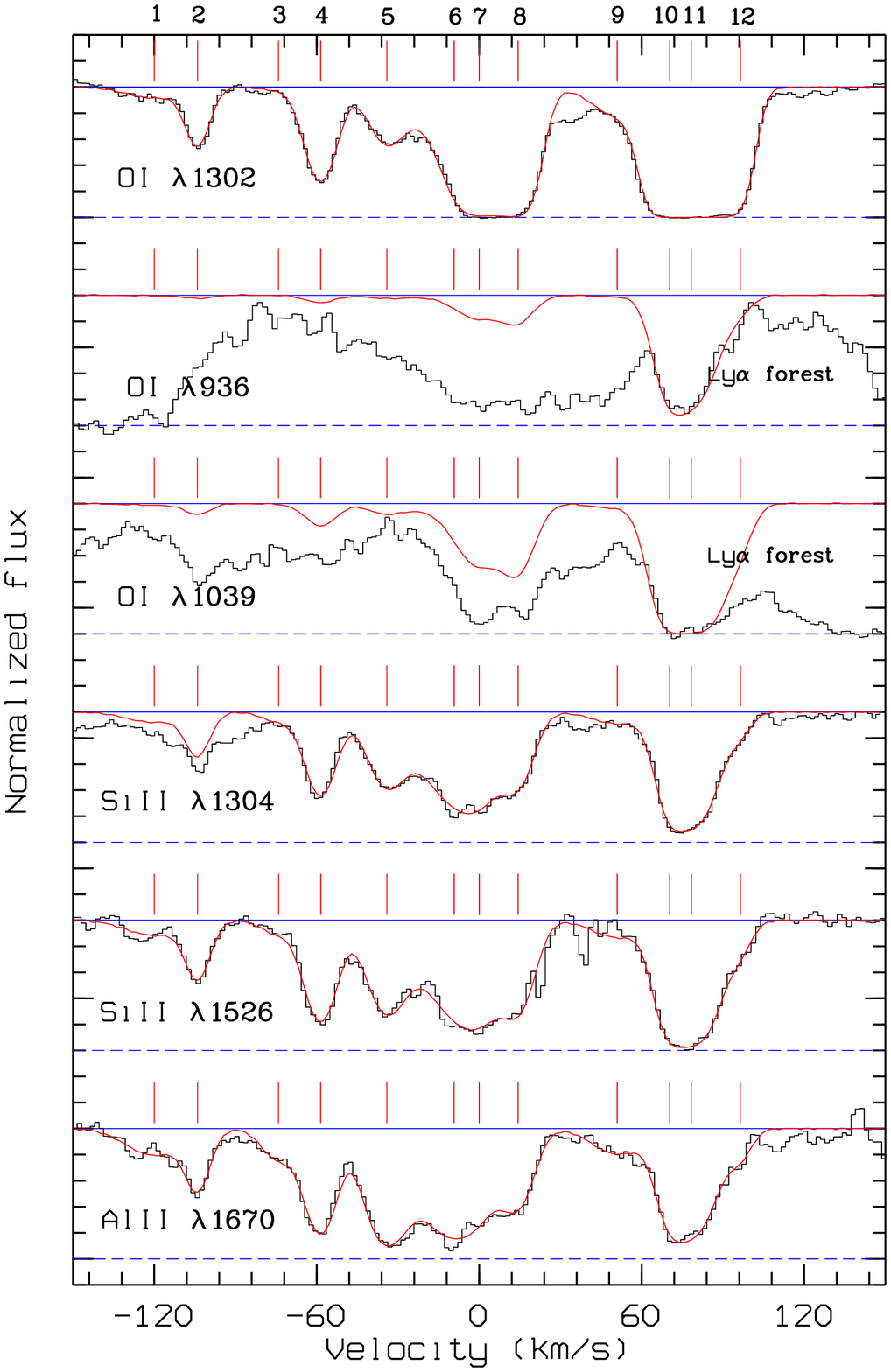,width=8.8cm,clip=}\quad
\psfig{figure=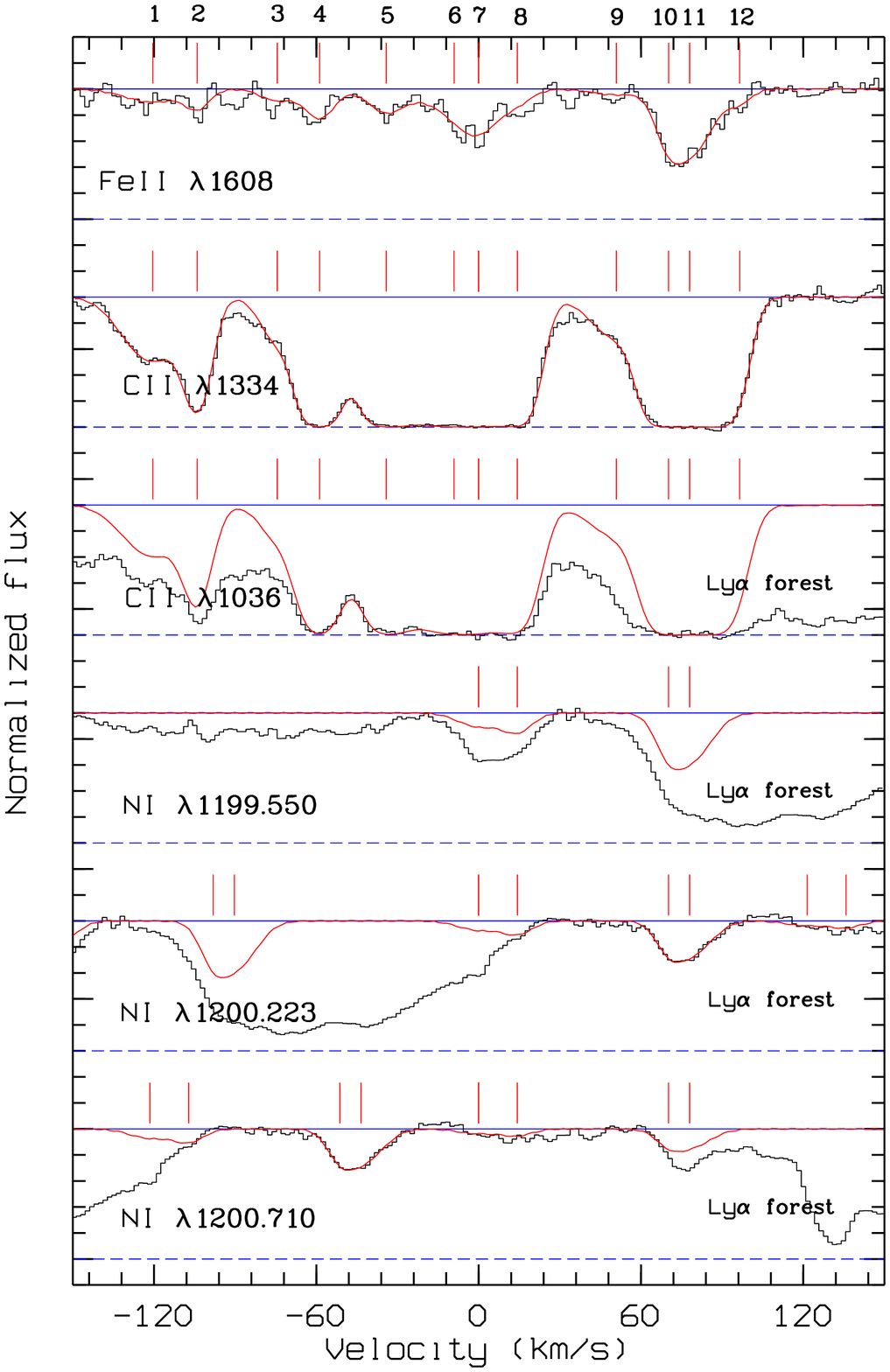,width=8.8cm,clip=}}
\end{center}
\caption[]{Absorption-line profiles of the low-ion transitions plotted against velocity for the DLA system at 
$z_{\rm abs} = 4.466$. The vertical scale goes from 0 to 1 for each plotted line transition. The zero velocity is 
fixed at $z = 4.466582$. The vertical bars mark the positions of the 12 velocity components given in Table~\ref{low-ion-tab}. 
The thin solid curve represents the best fit solution. The \ion{N}{i} triplet is partly contaminated by the Ly$\a$ forest 
absorptions, thus the fit was performed only for their 4 stronger components (7, 8, 10 and 11).}
\label{low-ion-gr}
\end{figure*}


The velocity profiles and the best fitting solutions of the low-ion transitions (C$^+$, N$^0$, O$^0$, Al$^+$, Si$^+$ and 
Fe$^+$) associated with the damped Ly$\a$ system at $z_{\rm abs} = 4.466$ are plotted in Fig.~\ref{low-ion-gr}. The 
velocity centroids for the different components of the fit are marked by the short vertical lines. To obtain these fitting 
solutions, we assumed the same $b$-values in the same component of all ions (i.e. that macroturbulent motions dominate over 
thermal broadening) and the same redshift, and allowed for variations from ion to ion in column density only. We find that a 
minimum of twelve components was required to fit the low-ion transitions. The stronger components are the components 7, 8 
(at $z = 4.466582$, $z = 4.466844$) and the components 10, 11 (at $z = 4.467863$, $z = 4.468005$). Table~\ref{low-ion-tab} 
lists the redshift, the $b$-value and the column density of every velocity component of the fitting solution.

The velocity profiles and the best fitting solutions for the high-ion transitions (C$^{3+}$ and Si$^{3+}$ partly blended 
by the telluric absorption lines), which were obtained independently from the low-ion transitions, are given in 
Fig.~\ref{high-ion-gr} and in Table~\ref{high-ion-tab}. A total of fourteen components were required to fit the high-ion 
transitions, with the dominant one lying at $z = 4.465953$ (component 8).

\subsection{Hydrogen content}

The hydrogen column density was mainly estimated from the fit of the Ly$\a$ damping profile. The main 
difficulty of the Ly$\a$ profile fitting in this DLA system, as is often the case at high redshift, is 
the strong contamination by Ly$\a$ forest lines on both wings of the damped Ly$\a$ line. The Ly$\beta$ and Ly$\gamma$ 
lines are heavily blended and thus unusable for a hydrogen column density determination, unlike the Ly$\delta$ line.
 
\begin{figure}[!]
\begin{center}
\psfig{figure=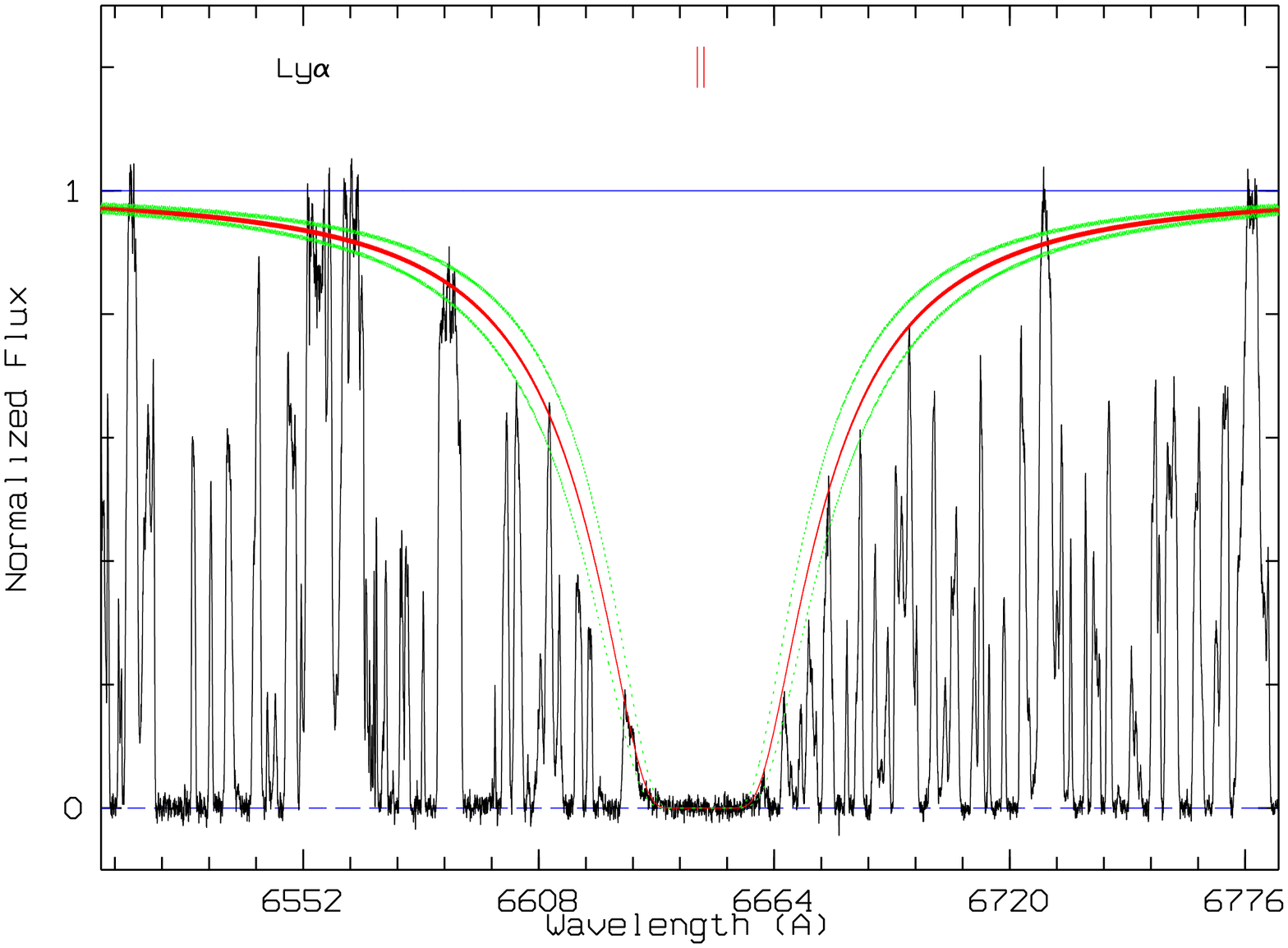,width=8.3cm,angle=-90,clip=}
\vspace{3mm}
\psfig{figure=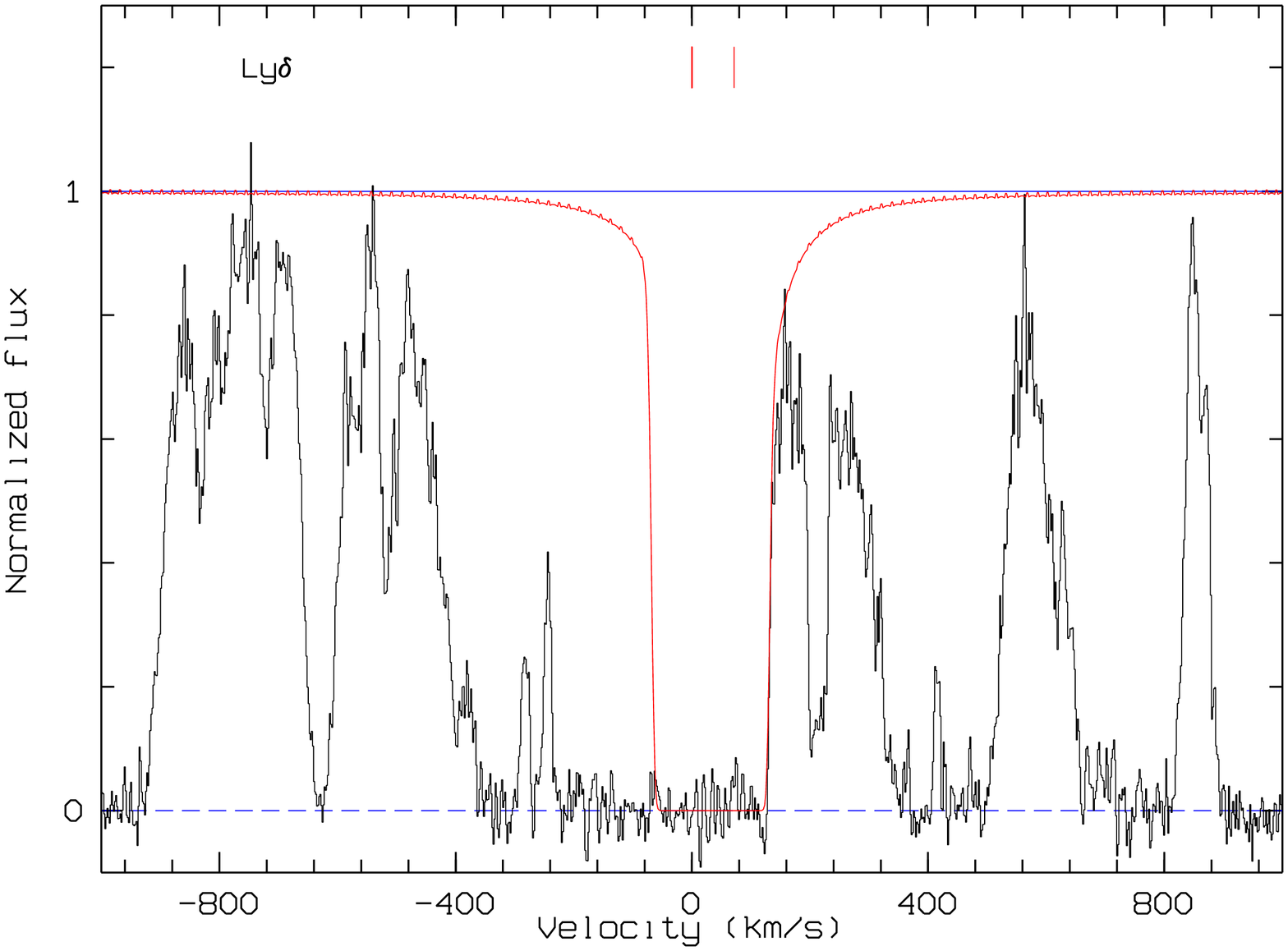,width=8.3cm,angle=-90}
\end{center}
\caption[]{Damped Ly$\a$ and Ly$\delta$ absorption profiles plotted against vacuum heliocentric wavelength ({\em top}) and 
velocity ({\em bottom}) for the DLA system at $z_{\rm abs} = 4.466$. The vertical bars mark the positions of 2 stronger 
components (7 and 11) at $z = 4.466582$ and $z = 4.468005$. The thin solid curve represents the best fit of the damping 
profile with $N$(\ion{H}{i})$= 1.07\times 10^{20}$ cm$^{-2}$, $b = 20.0$ km s$^{-1}$ and 
$N$(\ion{H}{i})$= 3.63\times 10^{20}$ cm$^{-2}$, $b = 16.5$ km s$^{-1}$ (respectively). The light dotted curves are damping 
profiles with $N$(\ion{H}{i})$= (0.86,2.90)\times 10^{20}$ cm$^{-2}$ and $N$(\ion{H}{i})$= (1.28,4.36)\times 10^{20}$ 
cm$^{-2}$ (respectively).}
\label{Lya}
\end{figure}

Fits provided by the Ly$\a$ and Ly$\delta$ lines by taking into account only one of the four stronger components of the 
low-ion fits (component 7, nearly centered on the Ly$\a$ and Ly$\delta$ profiles), yield a satisfactory result only to the 
blue side of the Ly$\a$ profile (the Ly$\delta$ blue side profile is contaminated by the Ly$\a$ forest absorptions). However, 
they leave an ``empty space'' on the red side on each of these two line profiles, which can be filled by adding the 
contribution of the second dominant component, the component 11 (see Fig.~\ref{Lya}).

The corresponding hydrogen column densities and $b$-values are $\log N$(\ion{H}{i}) = 20.03, $b$ = 20.0 km s$^{-1}$ for the 
component 7 at $z=4.466582$ and $\log N$(\ion{H}{i}) = 20.56, $b$ = 16.5 km s$^{-1}$ for the component 11 at $z=4.468005$. 
The $b$-value of the component 11 is constrained by the blue side of the Ly$\delta$ line. The resulting total hydrogen column 
density is $\log N$(\ion{H}{i}) $= 20.67 \pm 0.09$.

The hydrogen column density value determined in the DLA system at $z_{\rm abs} = 4.466$ is consistent with the lack of 
high neutral hydrogen column density DLA systems towards the highest redshifts (Storrie-Lombardi \& Wolfe 2000). 
Indeed, in the sample described in the latter reference, 6 DLA absorbers out of 10 at $3\leq z_{\rm abs} \leq 3.5$ have 
$\log N$(\ion{H}{i}) $>21$, whereas the statistics drops to 1 out of 12 at $z_{\rm abs} >3.5$.

\subsection{Metal abundances}

The ionic column densities and the relative metal abundances of C, N, O, Al, Si, S, Fe and Ni measured for 
this DLA system at $z_{\rm abs} = 4.466$ are summarized in Table~\ref{col-den}. We give the total ionic column 
densities obtained by adding up all the components derived from the fits and their corresponding errors. The errors 
reflect the data quality and the errors on the $b$-values. The relative metal abundances are estimated by assuming 
that the neutral and singly ionized species are associated with the neutral phase from which the \ion{H}{i} column 
density originates and consequently do not require ionization corrections (Viegas 1995, Lu et al. 1995). Upper limits 
of detection are computed under the optically thin case approximation by considering the contributions of the four 
stronger components (7, 8, 10 and 11) not detected at 4 $\sigma$. When the absorption lines are saturated, we give 
a lower limit.

The difficulty in determining accurate \ion{O}{i} column densities is due to the fact that the only strong transition 
available redwards of the Ly$\a$ emission is the \ion{O}{i}$\l$1302 transition which is usually saturated in DLA systems. 
This transition is also saturated in this system at $z_{\rm abs} = 4.466$ (see Fig.~\ref{low-ion-gr}), and thus allows to 
derive only a lower limit to the total $N$(\ion{O}{i}). We have looked for other oxygen lines in the Ly$\a$ forest namely, 
\ion{O}{i}$\l$921.857, $\l$929.517, $\l$936.629, $\l$948.685, $\l$950.885, $\l$988.773 and $\l$1039.230. Most of 
these lines are heavily contaminated by the Ly$\a$ forest absorption lines and cannot be used for any column density 
measurement. However, as we can see in Fig.~\ref{low-ion-gr}, the \ion{O}{i}$\l$936.629 and $\l$1039.230 lines put a 
stringent upper limit on the \ion{O}{i} column densities of the two strong components 10 and 11, which are saturated in the 
\ion{O}{i}$\l$1302 line. This plays a crucial role in favour of the accuracy of the determined $N$(\ion{O}{i}) column density 
value, since the measured upper limits are very close to the lower limit obtained from the \ion{O}{i}$\l$1302 line. Then, the 
only remaining saturated components in the \ion{O}{i}$\l$1302 line are the two other strong components 7 and 8 for which the 
other \ion{O}{i} lines do not provide any constraint on their column densities. Nevertheless, the error made 
on the column densities of these two components has a negligible effect on the total $N$(\ion{O}{i}) column density: 
if we suppose a 40\% underestimation of the column densities of the components 7 and 8 (which has an
important impact on the fit quality of the \ion{O}{i}$\l$1302 line, owing to the very high signal-to-noise ratio in 
this region of the spectrum), it implies a variation of the total $\log N$(\ion{O}{i}) of 0.02 only, a variation which 
is fully taken into account by the adopted error bar of 0.17 on $\log N$(\ion{O}{i}). We conclude that the \ion{O}{i} 
column density is constrained in the range $15.74 < \log N(\textrm{\ion{O}{i}}) < 16.08$.

Accurate determinations of the carbon column densities present the same difficulty as for oxygen. The 
\ion{C}{ii}$\l$1334 line redwards of the Ly$\a$ emission is heavily saturated and provides only a lower limit to 
$N$(\ion{C}{ii}), and the \ion{C}{ii}$\l$1036 line is blended with Ly$\a$ forest absorptions (Fig.~\ref{low-ion-gr}). 
The \ion{C}{ii}$^* \l$1335 metastable line is not detected, and thus provides only an upper limit for the \ion{C}{ii}$^*$ 
column density.


\begin{table*}[!]
\begin{center}
\caption{Fits for the DLA system at $z_{\rm abs} = 4.466$ - High-ions} 
\label{high-ion-tab}
\begin{tabular}{l c c l c | l c c l c}
\hline\hline
Comp & $z_{\rm abs}$ & $b$    & Ident & $\log N$    & Comp & $z_{\rm abs}$ & $b$    & Ident & $\log N$  \\
     &           & [km s$^{-1}$] &       & [cm$^{-2}$] &      &           & [km s$^{-1}$] &       & [cm$^{-2}$]  
\smallskip
\\     
\hline
1... & 4.463207 & 12.9 $\pm$ 6.0  & \ion{C}{iv} & 12.61 $\pm$ 0.10 & 8...  & 4.465953 & 11.8 $\pm$ 2.2   & \ion{C}{iv} & 13.72 $\pm$ 0.12 \\
     &          &                 & \ion{Si}{iv}& 12.30 $\pm$ 0.14 &       &          &                  & \ion{Si}{iv}& 13.58 $\pm$ 0.22\\
2... & 4.463557 & 12.1 $\pm$ 4.0  & \ion{C}{iv} & 12.93 $\pm$ 0.05 & 9...  & 4.466476 & 10.8 $\pm$ 1.5   & \ion{C}{iv} & 13.26 $\pm$ 0.03\\
     &          &                 & \ion{Si}{iv}& 12.62 $\pm$ 0.08 &       &          &                  & \ion{Si}{iv}& 13.24 $\pm$ 0.06\\
3... & 4.463931 & 11.3 $\pm$ 2.2  & \ion{C}{iv} & 13.13 $\pm$ 0.03 & 10... & 4.466779 & 12.7 $\pm$ 5.4   & \ion{C}{iv} & 12.78 $\pm$ 0.08\\
     &          &                 & \ion{Si}{iv}& 12.79 $\pm$ 0.05 &       &          &                  & \ion{Si}{iv}& 12.53 $\pm$ 0.06\\
4... & 4.464331 &  4.7 $\pm$ 1.4  & \ion{C}{iv} & 12.90 $\pm$ 0.05 & 11... & 4.467173 & 19.0 $\pm$ 6.9   & \ion{C}{iv} & 13.31 $\pm$ 0.03\\
     &          &                 & \ion{Si}{iv}& 12.79 $\pm$ 0.05 &       &          &                  & \ion{Si}{iv}& 13.14 $\pm$ 0.05\\
5... & 4.464525 & 17.8 $\pm$ 6.1  & \ion{C}{iv} & 13.09 $\pm$ 0.05 & 12... & 4.467606 & 10.3 $\pm$ 1.9   & \ion{C}{iv} & 13.13 $\pm$ 0.03\\
     &          &                 & \ion{Si}{iv}& 12.94 $\pm$ 0.05 &       &          &                  & \ion{Si}{iv}& 12.98 $\pm$ 0.04\\
6... & 4.465333 & 19.8 $\pm$ 4.7  & \ion{C}{iv} & 13.27 $\pm$ 0.03 & 13... & 4.467994 &  5.1 $\pm$ 2.6   & \ion{C}{iv} & 12.54 $\pm$ 0.07\\
     &          &                 & \ion{Si}{iv}& 12.89 $\pm$ 0.15 &       &          &                  & \ion{Si}{iv}& 12.45 $\pm$ 0.04\\
7... & 4.465565 &  4.0 $\pm$ 1.8  & \ion{C}{iv} & 12.72 $\pm$ 0.07 & 14... & 4.468213 &  4.3 $\pm$ 3.5   & \ion{C}{iv} & 12.42 $\pm$ 0.09\\
     &          &                 & \ion{Si}{iv}& 12.05 $\pm$ 0.21 &       &          &                  & \ion{Si}{iv}& 12.47 $\pm$ 0.04\\
\hline 
\end{tabular}
\end{center}
\end{table*}


\begin{figure}[!]
\begin{center}
\mbox{\psfig{figure=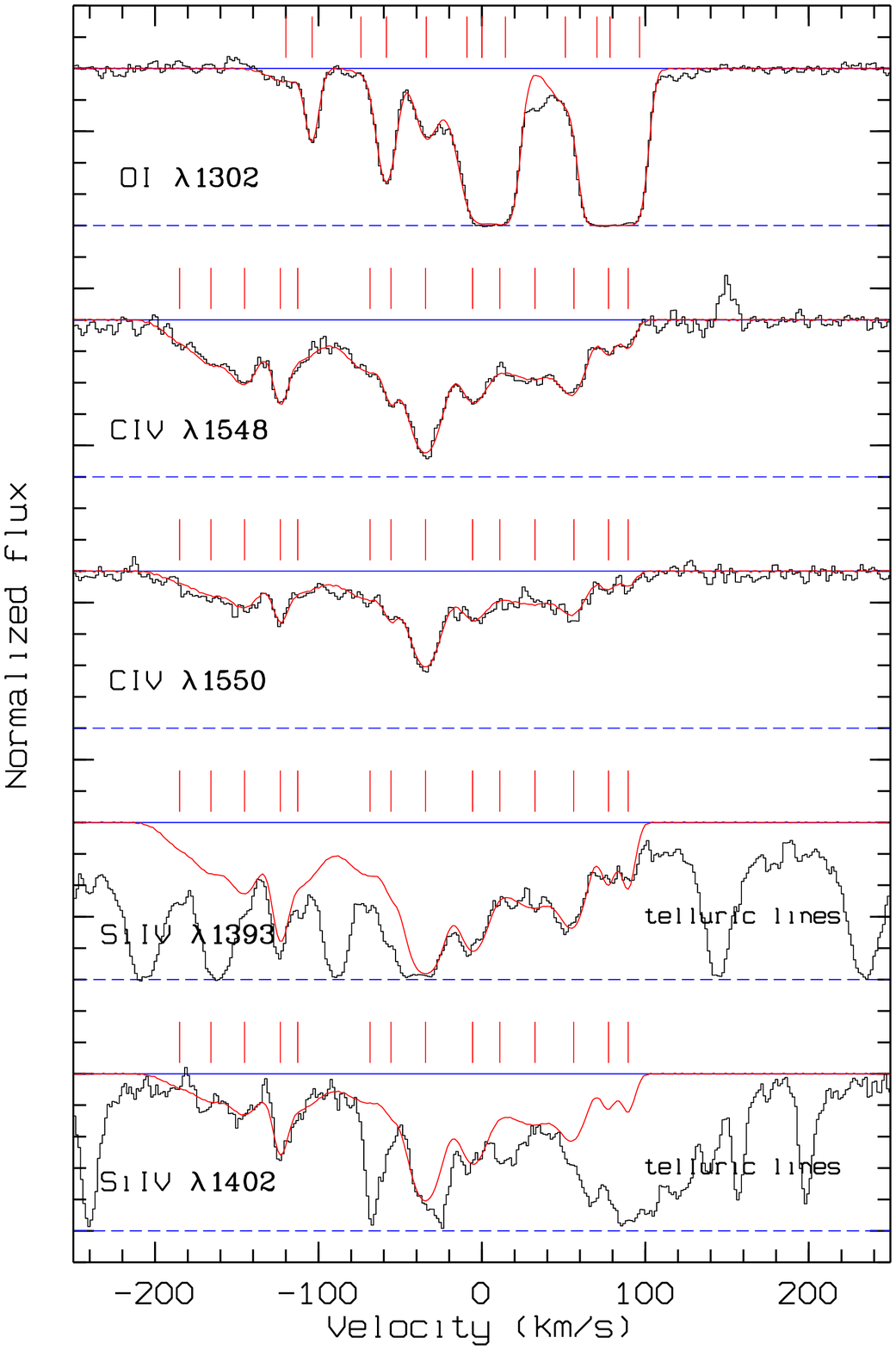,width=7.6cm,clip=}}
\end{center}
\caption[]{Absorption-line profiles of the high-ion transitions plotted against velocity for the DLA system at 
$z_{\rm abs} = 4.466$. The vertical scale goes from 0 to 1 for each plotted line transition. The zero velocity is 
fixed at $z = 4.466582$. The vertical bars mark the positions of the 14 velocity components as given in 
Table~\ref{high-ion-tab}. The thin solid curve represents the best fit solution. The \ion{Si}{iv}$\l\l$1393,1402 lines 
are partly blended with telluric absorption lines. The \ion{O}{i} profile is plotted in order to facilitate the comparison 
between low- and high-ion profiles.}
\label{high-ion-gr}
\end{figure}


Silicon has been derived from the \ion{Si}{ii}$\l$1304 and $\l$1526 lines and Al from the \ion{Al}{ii}$\l$1670 line. The 
\ion{Ni}{ii}$\l$1317 and $\l$1370 lines are not detected, and thus provide only an upper limit for the Ni column density. 
The iron abundance is obtained from the \ion{Fe}{ii}$\l$1608 line: 
[Fe/H]\footnote{[X/H] $\equiv \log$[$N$(X)/$N$(H)]$_{\scriptsize{\textrm{DLA}}}$ $- \log$[$N$(X)/$N$(H)]$_{\odot}$.} 
$= -1.97\pm 0.19$. Zn is a useful metal content indicator because it is undepleted in the interstellar medium (Pettini et 
al. 1997). Zn abundance measurements are however not possible in this DLA system, since the doublet \ion{Zn}{ii}$\l$2026 and 
$\l$2062 fall outside the working range of the spectrograph at an observed wavelength $\l > 11000$ \AA.

The \ion{N}{i} triplet near 1200 \AA\ is clearly detected (see Fig.~\ref{low-ion-gr}) in the Ly$\a$ forest, 
although only the four stronger components (7, 8, 10 and 11) are free of the Ly$\a$ forest absorption contamination. 
By fitting these four components and by adding up their measured column densities, we derived a satisfactory value 
of the $N$(\ion{N}{i}) column density (the contribution of the other eight components to the total column density is 
negligible compared to the contribution of these four dominant components). The \ion{S}{ii} triplet near 1253 \AA\ 
is also detected in the Ly$\a$ forest, but it is strongly contaminated by the Ly$\a$ forest absorptions. By fitting 
the four dominant components (7, 8, 10 and 11), an upper limit to the $N$(\ion{S}{ii}) column density has been derived.


\section{Relative abundances}

\subsection{Metallicity and dust content}

The metallicity [Fe/H] $= -1.97\pm 0.19$ of the most distant DLA system known to the present date at $z_{\rm abs} = 4.466$, 
observed when the age of the universe was only 1.3\footnote{Assuming H$_0 = 60$ km s$^{-1}$ Mpc$^{-1}$, $\Omega_0 = 0.4$, 
$\Lambda_0 = 0.6$.} Gyr, is low but still a factor of two higher than the most 
metal-poor Galactic stars. It is not even the lowest metallicity level of all known DLA systems. Several systems have 
metallicities much lower than that and some of them in fact do lie at $z_{\rm abs} < 3$. The lowest abundances known to 
date, [Fe/H] $< -2.6$, have been measured at $z_{\rm abs} = 2.076$ toward Q 2206$-$199 (Prochaska \& Wolfe 1997a), at 
$z_{\rm abs} = 2.618$ toward Q 0913$+$072 (Ledoux et al. 1998) and at $z_{\rm abs} = 4.203$ toward BR 0951$-$04 
(Prochaska \& Wolfe 1999). The [Fe/H] $= -1.97\pm 0.19$ value of this system indeed shows that the very young object 
($\leq 1.3$ Gyr) responsible for the DLA absorber has already experienced a significant metal production. 

Prochaska \& Wolfe (2000) have recently discussed the high-quality measurements of the Fe abundance in 39 DLA systems. 
They find a moderate trend (with a large scatter) of decreasing [Fe/H] for individual absorbers from $z_{\rm abs} = 1.5$ 
to $4.4$. The $N$(\ion{H}{i})-weighted mean metallicity exhibits on the contrary a minimal evolution over the same redshift 
interval. Our result at $z_{\rm abs} = 4.466$ is consistent with this recent metallicity evolution review, since the 
metallicity found, [Fe/H] $= -1.97\pm 0.19$, is representative of the unweighted mean [Fe/H] metallicity for their 
high-redshift sub-sample, $\langle \textrm{[Fe/H]}\rangle_{z_{high} = [3;4.4\lbrack}$ $= -1.83$. 

The estimated upper limit of the nickel abundance [Ni/H] $< -2.35$ gives [Ni/Fe] $< -0.41$. This upper limit is 
significantly lower than the [Ni/Fe] values found in Galactic metal-poor stars which are nearly solar down to very 
low metallicities (Goswami \& Prantzos 2000). This underabundance of [Ni/Fe], measured despite the use 
of the updated \ion{Ni}{ii} $f$-value from Zsarg\' o \& Ferderman (1998), may be interpreted at a first glance 
as the result of differential depletion of Ni and Fe from the gas phase onto dust grains by analogy to what is 
observed in the Galactic ISM with Ni being more heavily depleted than Fe (Savage \& Sembach 1996). However, 
this interpretation has to be considered with caution, since the measurement of Ni in the dust-free DLA system 
toward Q 0000$-$2620 by Molaro et al. (2000) shows also a significant underabundance of [Ni/Fe], [Ni/H] $= -2.27$ for 
a metallicity of [Fe/H] $= -2.04$. Thus, the [Ni/Fe] ratio is not a good indicator of dust content in DLA systems.

The best constraint on dust content we have for the DLA system at $z_{\rm abs} = 4.466$ comes from the nearly solar 
[O/Si] ratio of $-0.09\pm 0.18$. It is suggestive of limited amounts of dust, since O is a non-refractory element and a 
depletion of Si onto dust grains would imply an intrinsic [O/Si] ratio much too undersolar. The derived [O/Si] ratio is also 
in excellent agreement with the usual assumption of nearly solar [O/$\a$] ratios in DLA systems (e.g.  Lu et al. 1998). 
Limited amounts of dust in this DLA system allow a study of relative abundances exempt from strong uncertainties deriving 
from the unknown fraction of elements which condenses onto dust grains.


\begin{table}[!]
\begin{center}
\caption{Ionic column densities and relative abundances for the DLA system at $z_{\rm abs} = 4.466$} 
\label{col-den}
\begin{tabular}{l c c c c c}
\hline\hline
Ion             & $\l_{rest}$  & $f$         & $\log N$$^e$      & [X/H]$^f$ 
\smallskip
\\     
\hline
\ion{H}{i}..... & 1215.670$^a$ & 0.41640$^a$ & 20.67 $\pm$ 0.09  & ... \\
\ion{C}{ii}.... & 1334.532$^a$ & 0.12780$^a$ & $>$ 15.59	 & $>$ -1.63 \\
\ion{C}{ii}$^*$..&1335.708$^a$ & 0.11490$^a$ & $<$ 13.59	 & ... \\
\ion{C}{iv}...  & 1548.195$^a$ & 0.19080$^a$ & 14.27 $\pm$ 0.06  & ... \\
                & 1550.770$^a$ & 0.09522$^a$ & ...		 & ... \\
\ion{N}{i}..... & 1199.550$^a$ & 0.13280$^a$ & 13.57 $\pm$ 0.12  & -3.07 $\pm$ 0.15 \\
                & 1200.223$^a$ & 0.08849$^a$ & ...		 & ... \\
                & 1200.710$^a$ & 0.04423$^a$ & ...		 & ... \\
\ion{O}{i}..... & 1302.168$^a$ & 0.04887$^a$ & 15.91 $\pm$ 0.17  & -1.63 $\pm$ 0.19 \\
                & 1039.230$^a$ & 0.00920$^a$ & ...		 & ...\\
		&  936.629$^a$ & 0.00314$^a$ & ...		 & ... \\
\ion{Al}{ii}... & 1670.787$^a$ & 1.83300$^a$ & 13.36 $\pm$ 0.06  & -1.79 $\pm$ 0.11 \\
\ion{Si}{ii}... & 1304.370$^b$ & 0.08600$^b$ & 14.68 $\pm$ 0.07  & -1.54 $\pm$ 0.11  \\
                & 1526.707$^b$ & 0.11000$^b$ & ...		 & ...\\
\ion{Si}{iv}... & 1393.755$^a$ & 0.51400$^a$ & 14.09 $\pm$ 0.12  & ...\\
                & 1402.770$^a$ & 0.25530$^a$ & ...		 & ...\\
\ion{S}{ii}.... & 1250.584$^a$ & 0.00545$^a$ & $<$ 15.46	 & $<$ -0.41 \\
                & 1253.811$^a$ & 0.01088$^a$ & ...		 & ... \\
		& 1259.519$^a$ & 0.01624$^a$ & ...		 & ...  \\
\ion{Fe}{ii}... & 1608.451$^c$ & 0.06196$^c$ & 14.21 $\pm$ 0.17  & -1.97 $\pm$ 0.19 \\
\ion{Ni}{ii}... & 1314.217$^d$ & 0.14580$^d$ & $<$ 12.60	 & $<$ -2.35 \\
                & 1370.132$^d$ & 0.14400$^d$ & $<$ 12.57	 & ... \\
\hline 
\end{tabular}
\begin{minipage}{160mm}
\smallskip
$^a$ From Morton 1991.\\
$^b$ From Spitzer \& Fitzpatrick 1993.\\
$^c$ From Cardelli \& Savage 1995.\\
$^d$ From Zsarg\' o \& Federman 1998. \\
$^e$ Column density from line profile fitting. Upper limits are 4 $\sigma$.\\
$^f$ Abundance relative to the solar value of Grevesse et al. 1996.\\
\end{minipage}
\end{center}
\end{table}
 
 
\subsection{$\a$-element to Fe abundances}\label{a-to-Fe}

The [$\a$/Fe] abundance ratio is a good indicator of the chemical evolution history which can be used to understand the 
nature of galaxies. In the early stages of the chemical evolution of galaxies the abundances are likely dominated by Type II 
supernovae (SNe) products rich in $\a$-elements created within $< 2\times 10^7$ yr. The Type I SNe products, the 
Fe-peak elements, entered into the game only after $> 10^8 -10^9$ yr. Thus, in the early stages the
$\a$-elements are expected to be more abundant than the Fe-peak elements, and the ratio [$\a$/Fe] should decline later on 
during evolution. The exact timing of the decline depends on both the star formation rate and the initial mass function. 
A galaxy which turns most of its gas into stars within $10^9$ yr would maintain an enhanced [$\a$/Fe] 
ratio while [Fe/H] grows to high values (Fuhrmann 1998). At the other extreme, in a galaxy with a small star formation rate 
or with bursts separated by quiescent periods lasting more than $10^9$ yr, there would be time for [$\a$/Fe] to decline 
to at least solar values while [Fe/H] remains low (Pagel \& Tautvaisviene 1998).

Si is the most accessible $\a$-element in the DLA systems. However it is a refractory element, and due to the Si 
depletion coupled with a stronger Fe depletion onto dust grains (Savage \& Sembach 1996), we observe similar [Si/Fe] 
ratios in the Galactic interstellar medium (ISM) as in the Galactic metal-poor stars presumed to exhibit 
nucleosynthetic patterns typical of Type II supernovae. In the presence of dust, it is therefore difficult to determine 
the intrinsic [Si/Fe] enhancement. In the DLA system at $z_{\rm abs} = 4.466$, we derive [Si/H] $= -1.54\pm 0.11$ which 
implies [Si/Fe] $= 0.43\pm 0.18$, a relatively high $\a$ enhancement. At the metallicity of [Fe/H] $= -1.97\pm 0.19$, 
this [Si/Fe] value corresponds to the typical values of the Galactic metal-poor stars (Goswami \& Prantzos 2000), and in 
a dust-limited system it is suggestive of a Type II SN enrichment.

Oxygen is also a typical product of Type II SNe and shows an important enhancement in Galactic metal-poor stars, 
[O/Fe] $\simeq 0.35$ at [Fe/H] $= -2.00$ (McWilliam 1997). Claims for an even more extreme overabundance of O with 
[O/Fe] reaching 1.0 dex at [Fe/H] $= -3.0$ have recently been made by Israelian et al. (1998) and Boesgaard et al. (1999). 
The oxygen dust depletion is negligible (Savage \& Sembach 1996), therefore O is an excellent diagnostic tool 
of the [$\a$/Fe] abundance ratio. In the studied DLA system, we derive [O/H] $= -1.63\pm 0.19$, which implies 
[O/Fe] $= 0.34\pm 0.24$. This value is representative of the typical Type II SN enriched Galactic metal-poor star 
abundance pattern.

Another important diagnostic tool of the [$\a$/Fe] abundance ratio is sulphur. S also shows in Galactic
metal-poor stars the typical enhancement of $\a$-elements with [S/Fe] $\simeq 0.4/0.6$ (Fran\c cois 1988), and it is 
nearly undepleted from gas to dust (Savage \& Sembach 1996). We manage to derive only a non-significant upper limit 
[S/H] $< -0.41$, which does not provide any interesting constraint on the [S/Fe] ratio ([S/Fe] $< 1.56$).

We have finally at our disposal essentially two indicators of the $\a$-element abundances, Si and O. The [Si/Fe] 
$= 0.43\pm 0.18$ and [O/Fe] $= 0.34\pm 0.24$ ratios both show a significant [$\a$/Fe] enhancement due to the Type II SN 
enrichment similar to that of the Galactic stars with comparable metallicities. A [$\a$/Fe] enhancement due to dust 
depletion effects is ruled out by the presence of limited amounts of dust in this DLA system. We can thus suggest that 
the DLA system towards APM BR J0307$-$4945 is undergoing a chemical evolution enrichment similar to that expected in 
the initial phases of a massive galaxy like the Milky Way.

\subsection{Nitrogen abundance}

Nitrogen is a key element in understanding the evolution of galaxies with few star forming events, since it needs relatively 
long timescales as well as relatively high underlying metallicity to be produced (Matteucci et al. 1997). The reason is that 
N is believed to be produced principally in intermediate-mass stars, specifically those stars between 4 and 8 M$_{\odot}$ 
which undergo hot bottom burning and expel large amounts of primary N at low metallicities and secondary N at higher 
metallicities produced in the CNO cycle from C and O created in earlier generations of stars (Henry et al. 2000).

In our system the N abundance is [N/H] $= -3.07\pm 0.15$, which implies [N/O] $= -1.44\pm 0.21$ at [O/H] $= -1.63\pm 0.19$. 
This very low [N/O] ratio, much lower than the corresponding values in the Milky Way and in the \ion{H}{ii} regions, lies in 
the region limited by the tracks of primary and secondary N production (Vila-Costas \& Edmunds 1993), close to a pure 
secondary N production behaviour.

All studied DLA systems occupy the same delimited region, and show a large scatter with some [N/O] values consistent with a 
pure secondary behaviour and others requiring a primary production of N (Lu et al. 1998, Centur\' \i on et al. 1998). The 
different [N/O] values observed at a given [O/H] may be interpreted in terms of the delayed delivery of N with respect to 
O when star formation proceeds in bursts (Edmunds \& Pagel 1978). In this model, galaxies that have experienced a recent 
episode of star formation show low [N/O] ratios due to the quick production of O in massive stars ($\sim 6$ Myr for a 25
M$_{\odot}$ star). Galaxies that have been quiescent for a long period show high [N/O], N is released after approximatively 
250 Myr as a product of intermediate-mass stars and the [N/O] ratio increases while [O/H] remains constant. Therefore, the 
scatter comes about by observing DLA systems in various stages of O and N enrichment.

The derived very low [N/O] ratio may thus be explained as representing very early stages of N/O evolution when
intermediate-mass stars have not yet begun to release N, i.e. roughly 250 Myr or less after a star burst. It thus seems 
reasonable to expect also an overabundance of the $\a$-elements relative to Fe-peak elements, an overabundance which has 
been highlighted by the [Si/Fe] and [O/Fe] abundance ratios in Sect.~\ref{a-to-Fe}.


\section{Kinematics}

The metal line profiles of DLA absorbers in principle provide information in velocity space on the gas kinematics. Whether 
this information can be used as a diagnostic of motions on galactic scales and what motions indeed dominate the kinematics is 
still an open question. The kinds of motion one may expect to be at play in a gas-rich environment, and in proto-galaxies, 
are rotation of clouds in a disk-like structure, and radial infall and random motions of clouds in a halo. A realistic 
description may have to combine both effects as is probably the case in \ion{Mg}{ii} and Lyman-limit systems at intermediate 
redshifts (Charlton \& Churchill 1996). In DLA systems, systematic high-resolution studies of low-ionization lines have 
established that the profiles are often asymmetric with the strongest absorption lying at an edge of the profile 
(Prochaska \& Wolfe 1997b, 1998). This was first interpreted as the signature of large, fast-rotating thick disks although 
the current body of data is also consistent with relative motions of clouds with limited rotation (Haehnelt et al. 1998, 
Ledoux et al. 1998, McDonald \& Miralda-Escud\'e 1999).

In the DLA absorber towards APM BR J0307$-$4945, the shape of the unsaturated \ion{Fe}{ii}$\l$1608 line profile is strongly 
suggestive of edge-leading asymmetry in the sense that the strongest absorption itself is located at an edge of the profile 
(see Fig.~\ref{low-ion-gr}). The signal-to-noise ratio is not high enough in this spectral region, however, to decide whether 
the entire profile is asymmetric, with steadily decreasing optical depth with decreasing wavelength. Stronger transition 
lines from different elements including \ion{Al}{ii} and \ion{O}{i}, the latter being a good tracer of the neutral phase, 
indeed display huge variations of optical depths along their profiles. In particular, whereas the low-ion profiles extend 
over 240 km s$^{-1}$, they have relatively small or negligible optical depths in a spectral region 25 km s$^{-1}$ wide just 
blueward of the strong components at the red edge of the profiles. Moreover, the best fit model described in 
Sect.~\ref{column-densities} indeed indicates that about 80\% of the metals are distributed in two sub-systems of 
comparable importance, separated by $\sim 70$ km s$^{-1}$ and spanning only about 25 km s$^{-1}$ each (components 6, 7, 8 on 
the one hand, and 10, 11, 12 on the other hand). The kinematics of the absorber can therefore also be understood as having a 
main double peaked structure. This is expected, for instance, during a merging event involving two proto-galactic clumps 
surrounded by tidal material and radially infalling clouds; this possibility is strongly supported by the results of numerical 
N-body/hydrodynamic simulations (see Haehnelt et al. 1998, 2000).

The comparison of the relative strength and velocity distribution of line profiles arising from different phases also 
provides information on the ionization structure and spatial distribution of the absorbing gas. In the DLA system under 
study, the shapes of the \ion{C}{iv} and \ion{Si}{iv} profiles are similar and globally symmetric with the strongest 
absorption located at the center of the profile; they also clearly differ from the low ions. The total velocity broadening 
is larger for the high-ion (305 km s$^{-1}$) than it is for the low-ion profiles (240 km s$^{-1}$). However the two different 
phases are related, it has been shown from a statistical point of view that there is a trend for the total velocity 
broadenings of \ion{C}{iv} and \ion{Fe}{ii} to be correlated (Ledoux et al. 1998; see also Wolfe \& Prochaska 2000). Besides, 
it is apparent in Fig.~\ref{high-ion-gr} that, while there is a velocity interval common to both phases, the high-ion 
profiles extend much farther into the blue. The DLA absorber towards APM BR J0307$-$4945 may thus be a good illustration of 
a scenario in which ionized gas is flowing out with velocities of up to $\sim -300$ km s$^{-1}$ relative to the bulk of 
one of two main neutral clumps.


\section{Conclusion}

Observations with the new UVES echelle spectrograph at the 8.2m VLT Kueyen telescope provided the first high-quality 
spectra of the distant APM BR J0307$-$4945 quasar at $z_{\rm em} = 4.73$. We focused our analysis on the metal abundances 
of a damped Ly$\a$ system at $z_{\rm abs}=4.466$, the most distant DLA system known to the present date, observed when the 
age of the universe was only 1.3 Gyr. It has a hydrogen column density of $N$(\ion{H}{i}) $= (4.68\pm 0.97)\cdot10^{20}$ 
cm$^{-2}$, and shows complex low- and high-ion line profiles spanning $\approx 240$ and $300$ km s$^{-1}$ in velocity 
space respectively. By fitting Voigt profiles to the observed absorption lines and by neglecting ionization corrections, 
we obtained [N/H] $= -3.07\pm 0.15$, [O/H] $= -1.63\pm 0.19$, [Al/H] $= -1.79\pm 0.11$, [Si/H] $= -1.54\pm 0.11$, 
[Fe/H] $= -1.97\pm 0.19$, and we placed a lower limit on the abundance of C, [C/H] $> -1.63$ and an upper limit on the 
abundance of Ni, [Ni/H] $< -2.35$.

The Fe abundance, [Fe/H] $= -1.97\pm 0.19$, is representative of the mean [Fe/H] metallicity (unweighted by the \ion{H}{i} 
column density) for the high-redshift sub-sample of Prochaska \& Wolfe (2000), and is thus consistent with a moderate 
metallicity decrease towards high redshifts. This metallicity, $\sim 1/90$ solar, also shows that the very young object 
($\leq 1.3$ Gyr) responsible for the DLA absorber has already experienced a significant metal enrichment.

The relative [$\a$/Fe] abundance ratios resemble those expected in the early phases of evolution of a massive galaxy such 
as our own Galaxy. The [Si/Fe] $= 0.43\pm 0.18$ and [O/Fe] $= 0.34\pm 0.24$ ratios indeed show both a significant [$\a$/Fe] 
enhancement due to Type II SN enrichment similar to the one of the Galactic stars with comparable metallicities, in the 
presence of limited amounts of dust as inferred by the nearly solar [O/Si] ratio. The Type II SN enrichment is further 
supported by the very low [N/O] $= -1.44\pm 0.21$ ratio. 


\begin{acknowledgements}

We are indebted to all people involved in the conception, construction and commissioning of UVES for the high quality of 
the spectra obtained early in the operation of the instrument. We warmly thank J.X. Prochaska and J. Bergeron for 
useful comments on an earlier version of the paper and P. Bristow for carefully checking the english text. M.D.-Z. is 
supported by an ESO Studentship and the Swiss National Funds.

\end{acknowledgements}




\begin{table*}[!]
\begin{center}
\caption{Identified metal systems from absorption lines redwards of the Ly$\a$ emission for the quasar 
APM BR J0307$-$4945} 
\label{id}
\begin{tabular}{c c c c | c c c c}
\hline\hline
$\l _{vac}$ & W       & Ident & $z_{\rm abs}$ & $\l _{vac}$ & W       & Ident & $z_{\rm abs}$ \\
$[$\AA]     & $[$\AA] &       &           & [\AA]       & [\AA]   &       &            
\smallskip
\\ 
\hline
6954.33 & 2.550 $\pm$ 0.074 & \ion{C}{ii}$\l$1334  & 4.211 & 8067.64 & 3.508 $\pm$ 0.273 & \ion{C}{iv}$\l$1548  & 4.211 \\
6964.28 & 0.987 $\pm$ 0.039 & \ion{C}{ii}$\l$1334  & 4.218 & 8078.48 & 5.708 $\pm$ 0.305 & \ion{C}{iv}$\l$1548  & 4.218 \\
7012.54 & 0.071 $\pm$ 0.008 & \ion{Si}{iv}$\l$1393 & 4.031 & 8080.78 & 4.098 $\pm$ 0.428 & \ion{C}{iv}$\l$1550  & 4.211 \\
7107.39 & 0.968 $\pm$ 0.085 & \ion{C}{iv}$\l$1548  & 3.591 & 8091.92 & 3.269 $\pm$ 0.428 & \ion{C}{iv}$\l$1550  & 4.218 \\
7118.51 & 3.106 $\pm$ 0.151 & \ion{O}{i}$\l$1302   & 4.466 & 8113.90 & 0.080 $\pm$ 0.033 & \ion{Ca}{ii}$\l$3969 & 1.044 \\
7130.40 & 2.364 $\pm$ 0.119 & \ion{Si}{ii}$\l$1304 & 4.466 & 8267.33 & 0.267 $\pm$ 0.107 &   ?                  & \\
7139.90 & 0.288 $\pm$ 0.059 & \ion{Ni}{ii}$\l$1370 & 4.211 & 8302.34 & 0.282 $\pm$ 0.087 &                      & \\
7141.83 & 0.084 $\pm$ 0.017 &  ?                   &       & 8344.94 & 2.245 $\pm$ 0.316 & \ion{Si}{ii}$\l$1526 & 4.466 \\
7142.57 & 0.080 $\pm$ 0.016 &  ?                   &       & 8436.70 & 0.144 $\pm$ 0.072 &   ?                  & \\
7155.86 & 0.072 $\pm$ 0.023 &  ?                   &       & 8463.15 & 2.361 $\pm$ 0.304 & \ion{C}{iv}$\l$1548  & 4.466 \\
7167.68 & 0.050 $\pm$ 0.017 &  ?                   &       & 8477.34 & 1.484 $\pm$ 0.324 & \ion{C}{iv}$\l$1550  & 4.466 \\
7177.07 & 0.086 $\pm$ 0.027 &  ?                   &       & 8506.34 & 0.321 $\pm$ 0.041 & \ion{Fe}{ii}$\l$2344 & 2.629 \\
7264.03 & 3.505 $\pm$ 0.158 & \ion{Si}{iv}$\l$1393 & 4.211 & 8508.96 & 0.430 $\pm$ 0.052 & \ion{Fe}{ii}$\l$2344 & 2.630 \\
7273.05 & 3.538 $\pm$ 0.258 & \ion{Si}{iv}$\l$1393 & 4.218 & 8616.15 & 0.125 $\pm$ 0.041 & \ion{Fe}{ii}$\l$2374 & 2.629 \\
7283.88 & 0.177 $\pm$ 0.075 &                      &       & 8618.75 & 0.167 $\pm$ 0.042 & \ion{Fe}{ii}$\l$2374 & 2.630 \\
7294.96 & 4.562 $\pm$ 0.275 & \ion{C}{ii}$\l$1334  & 4.466 & 8646.29 & 0.641 $\pm$ 0.073 & \ion{Fe}{ii}$\l$2382 & 2.629 \\
7309.88 & 2.831 $\pm$ 0.178 & \ion{Si}{iv}$\l$1402 & 4.211 & 8648.86 & 0.820 $\pm$ 0.082 & \ion{Fe}{ii}$\l$2382 & 2.630 \\
7320.09 & 1.930 $\pm$ 0.262 & \ion{Si}{iv}$\l$1402 & 4.218 & 8650.44 & 0.388 $\pm$ 0.085 & \ion{C}{iv}$\l$1548  & 4.587 \\
7347.91 & 0.154 $\pm$ 0.043 & \ion{C}{iv}$\l$1548  & 3.746 & 8654.34 & 0.079 $\pm$ 0.055 & \ion{C}{iv}$\l$1548  & 4.590 \\
7360.07 & 0.053 $\pm$ 0.034 & \ion{C}{iv}$\l$1550  & 3.746 & 8662.13 & 0.067 $\pm$ 0.024 &                      & \\
7370.40 & 0.578 $\pm$ 0.051 & \ion{C}{iv}$\l$1548  & 3.760 & 8664.78 & 0.297 $\pm$ 0.093 & \ion{C}{iv}$\l$1550  & 4.587 \\
7382.67 & 0.379 $\pm$ 0.054 & \ion{C}{iv}$\l$1550  & 3.760 & 8668.66 & 0.090 $\pm$ 0.046 & \ion{C}{iv}$\l$1550  & 4.590 \\
7384.58 & 0.830 $\pm$ 0.085 & \ion{Fe}{ii}$\l$1608 & 3.591 & 8686.18 & 0.455 $\pm$ 0.127 & \ion{C}{iv}$\l$1548  & 4.610 \\
7470.32 & 0.711 $\pm$ 0.102 & \ion{C}{iv}$\l$1548  & 3.825 & 8700.80 & 0.200 $\pm$ 0.127 & \ion{C}{iv}$\l$1550  & 4.610 \\
7482.75 & 0.428 $\pm$ 0.099 & \ion{C}{iv}$\l$1550  & 3.825 & 8706.55 & 0.754 $\pm$ 0.199 & \ion{Al}{ii}$\l$1670 & 4.211 \\
7618.87 &  {\em blended$^*$}& \ion{Si}{iv}$\l$1393 & 4.466 & 8719.02 & 0.435 $\pm$ 0.234 & \ion{Al}{ii}$\l$1670 & 4.218 \\
7668.13 &  {\em blended$^*$}& \ion{Si}{iv}$\l$1402 & 4.466 & 8775.55 & 0.086 $\pm$ 0.047 &   ?                  & \\
7670.75 & 1.971 $\pm$ 0.128 & \ion{Al}{ii}$\l$1670 & 3.591 & 8791.73 & 0.906 $\pm$ 0.289 & \ion{Fe}{ii}$\l$1608 & 4.466 \\
7748.87 & 0.048 $\pm$ 0.032 &  ?                   &       & 8794.97 & 0.319 $\pm$ 0.053 &   ?                  & \\
7785.87 & 0.056 $\pm$ 0.024 &  ?                   &       & 9132.58 & 2.581 $\pm$ 0.366 & \ion{Al}{ii}$\l$1670 & 4.466 \\
7787.36 & 0.172 $\pm$ 0.056 & \ion{Si}{iv}$\l$1393 & 4.587 & 9152.38 & 0.065 $\pm$ 0.050 &   ?                  & \\
7789.51 & 0.133 $\pm$ 0.041 & \ion{C}{iv}$\l$1548  & 4.031 & 9197.92 & 0.098 $\pm$ 0.046 &   ?                  & \\
7791.05 & 0.061 $\pm$ 0.036 & \ion{Si}{iv}$\l$1393 & 4.590 & 9386.06 & 0.276 $\pm$ 0.078 & \ion{Fe}{ii}$\l$2586 & 2.629 \\
7802.54 & 0.079 $\pm$ 0.036 & \ion{C}{iv}$\l$1550  & 4.031 & 9389.16 & 0.829 $\pm$ 0.196 & \ion{Fe}{ii}$\l$2586 & 2.630 \\
7837.81 & 0.046 $\pm$ 0.025 & \ion{Si}{iv}$\l$1402 & 4.587 & 9416.03 & 0.080 $\pm$ 0.062 &   ?                  & \\
7841.41 & 0.031 $\pm$ 0.021 & \ion{Si}{iv}$\l$1402 & 4.590 & 9435.16 & 0.587 $\pm$ 0.088 & \ion{Fe}{ii}$\l$2600 & 2.629 \\
7955.78 & 0.714 $\pm$ 0.140 & \ion{Si}{ii}$\l$1526 & 4.211 & 9437.99 & 0.676 $\pm$ 0.121 & \ion{Fe}{ii}$\l$2600 & 2.630 \\
7967.11 & 0.420 $\pm$ 0.304 & \ion{Si}{ii}$\l$1526 & 4.218 & 9489.26 & 0.166 $\pm$ 0.095 &   & \\ 
8042.74 & 0.139 $\pm$ 0.046 & \ion{Ca}{ii}$\l$3934 & 1.044 &         &                   & &  \\
\hline
\end{tabular}
\begin{minipage}{160mm}
\smallskip
$^*$ Line profiles contaminated by telluric lines; ? Unidentified
\end{minipage}
\end{center}
\end{table*}


\end{document}